\documentclass[preprint,authoryear,12pt]{elsarticle}

\usepackage{graphicx}

\usepackage{amssymb}

\usepackage{lineno}
\usepackage{color}

\journal{Journal of Atmospheric and Solar-Terrestrial Physics}

\begin{document}

\begin{frontmatter}



\title{Forward modelling to determine the observational signatures
of white-light imaging and interplanetary scintillation for the
propagation of an interplanetary shock in the ecliptic plane}


 \author[label1]{Ming Xiong\corref{cor1}}
 \ead{mmx@aber.ac.uk}
 \cortext[cor1]{Corresponding Author:} 

 \author[label1]{A. R. Breen}
 \author[label1]{M. M. Bisi}
 \author[label2]{M. J. Owens}
 \author[label1]{R. A. Fallows}
 \author[label3]{G. D. Dorrian}
 \author[label4]{J. A. Davies}
 \author[label5]{P. Thomasson}

 \address[label1]{Aberystwyth University}
 \address[label2]{Reading University}
 \address[label3]{Queen's University Belfast}
 \address[label4]{STFC Rutherford-Appleton Laboratory}
 \address[label5]{Jodrell Bank Observatory, University of Manchester}


\author{}

\address{}

\begin{abstract}
Recent coordinated observations of interplanetary scintillation
(IPS) from the EISCAT, MERLIN, and STELab, and stereoscopic
white-light imaging from the two heliospheric imagers (HIs) onboard
the twin STEREO spacecraft are significant to continuously track the
propagation and evolution of solar eruptions throughout
interplanetary space. In order to obtain a better understanding of
the observational signatures in these two remote-sensing techniques,
the magnetohydrodynamics of the macro-scale interplanetary
disturbance and the radio-wave scattering of the micro-scale
electron-density fluctuation are coupled and investigated using a
newly-constructed multi-scale numerical model. This model is then
applied to a case of an interplanetary shock propagation within the
ecliptic plane. The shock could be nearly invisible to an HI, once
entering the Thomson-scattering sphere of the HI. The asymmetry in
the optical images between the western and eastern HIs suggests the
shock propagation off the Sun-Earth line. Meanwhile, an IPS signal,
strongly dependent on the local electron density, is insensitive to
the density cavity far downstream of the shock front. When this
cavity (or the shock nose) is cut through by an IPS ray-path, a
single speed component at the flank (or the nose) of the shock can
be recorded; when an IPS ray-path penetrates the sheath between the
shock nose and this cavity, two speed components at the sheath and
flank can be detected. Moreover, once a shock front touches an IPS
ray-path, the derived position and speed at the irregularity source
of this IPS signal, together with an assumption of a radial and
constant propagation of the shock, can be used to estimate the later
appearance of the shock front in the elongation of the HI field of
view. The results of synthetic measurements from forward modelling
are helpful in inferring the in-situ properties of coronal mass
ejection from real observational data via an inverse approach.

%
%
%

\end{abstract}

\begin{keyword}
Heliospheric Imaging, Interplanetary Scintillation, Multi-Scale
Modelling



\end{keyword}

\end{frontmatter}

\linenumbers

\section{Introduction}
\subsection{Interplanetary Space}
Interplanetary space can be considered as a transmission channel
connecting the Sun and the Earth, permeated with the ubiquitous
magnetized solar wind flow from the Sun. The solar wind is
inherently bimodal with fast solar wind from coronal holes and slow
solar wind above coronal streamers \citep{McComas2000}. As a
consequence of the root of the spiral interplanetary magnetic field
(IMF) on the rotating Sun, the interface between the fast and slow
solar winds at the same latitude gradually develops into a
co-rotating interaction region (CIR) with increasing heliocentric
distance. The ambient radial plasma flow in the co-rotating
inhomogeneous background structure is frequently interrupted by
coronal mass ejections (CMEs). A CME could undergo significant,
nonlinear, and irreversible evolution during its interplanetary
propagation, interacting with the ambient structured medium and
other CMEs. An individual CME could have its outer magnetic shell
stripped away to form a diffusive boundary layer by magnetic
reconnection \citep{Wei2006}, be significantly pushed at its rear
boundary by a CIR \citep{Dal_Lago2006}, and even be entrained by a
CIR \citep{Rouillard2009}. The coupling of multiple CMEs in the
Sun-Earth system could result in a complex ejecta
\citep{Burlaga2002} or a shock-penetrated magnetic cloud (MC)
\citep{Lepping1997}. Particularly, the compression effect
accompanying CMEs colliding can intensify the southward magnetic
field and is subsequently responsible for major geomagnetic storms
\citep{Burlaga1987}. In terms of space weather, interplanetary space
should be emphasized as a pivotal node in the development of the
solar-terrestrial causal process.

\subsection{Remote-Sensing Techniques}
The interplanetary space environment has been sampled frequently by
observing interplanetary scintillation (IPS) of radio waves,
beginning with the pioneering work of \citet{Hewish1964}. IPS is
essentially the intensity scintillation received at a terrestrial
radio antenna as a result of a drifting interference pattern across
the IPS ray-path, formed by density fluctuations in the solar wind.
The intensity variance on the time-scale of 0.1 $\sim$ 10 seconds
for the micro-scale density irregularities, with a characteristic
scale of tens to a few hundreds of kilometers, can be sequentially
recorded by two telescopes if the baseline between two telescopes is
basically parallel to the drifting speed of density irregularities
\citep{Bisi2006}. Long baselines can resolve multiple solar wind
streams crossing the IPS ray-path, which contribute to the integral
intensity towards the Earth from the different radio-scattering
layers along the IPS ray-path. The stream interface could be a shear
and sliding layer between a fast stream at the high latitude and a
slow stream at the ecliptic plane, or a compression layer within the
CIR, as the IPS observations reveal the velocity gradient and normal
scintillation level for the sliding layer, and an intermediate
velocity and enhanced scintillation level for the compression layer
\citep{Bisi2010}. The presence of a CME amid the background of
bimodal solar-wind streams \citep{Dorrian2008} can be identified
from the IPS signal as (1) a noticeable negative lobe in the
cross-correlation function (CCF), (2) a rapid variation of the solar
wind speed, the CCF shape, and the scintillation level on the
time-scale of hours or less. The poleward deflection of the ambient
solar wind ahead of a CME on 13 May 2005 \citep{Breen2008} and the
micro-structure of two CMEs merging on 16 May 2007
\citep{Dorrian2008} were reported by the IPS observations, using the
European Incoherent SCATter radar (EISCAT) in northern Scandinavia,
the Multi-Element Radio-Linked Interferometer Network (MERLIN) radio
telescopes in United Kingdom, and the Solar-Terrestrial Environment
Laboratory (STELab) in Japan. The capability of the traditional and
economical IPS technique to probe the inner heliosphere has been
significantly improved by the current advances of long baselines of
over 2000 km, the single observing frequency of around 1.4 GHz, and
dual-frequency observations of IPS \citep{Fallows2006}.

Modern heliospheric imaging and the traditional IPS techniques can
routinely cover the whole interplanetary space to fill in the
relative observation data gap between the comprehensively monitored
Sun and Earth. On the one hand, stereoscopic optical observations
using Heliospheric Imagers (HIs) \citep{Howard2008,Eyles2009} have
been successfully realized as a major milestone since the launch of
the twin Solar Terrestrial Relations Observatory (STEREO) spacecraft
in 2006 \citep{Kaiser2008}. Occupying solar orbits approximately at
1 AU, the STEREO A leads the Earth in the west and the STEREO B lags
the Earth in the east. Both STEREO A and B are separated from the
Earth by $22.5^\circ$ per year. Onboard either STEREO spacecraft,
each HI instrument comprises two cameras of HI-1 and HI-2, whose
optical axes lie in the ecliptic plane. Seen from a spacecraft, an
angular distance between the Sun and a target is defined by an
elongation. The elongation coverage is $4^\circ \sim 24^\circ$ for
the HI-1 and $18.7^\circ \sim 88.7^\circ$ for the HI-2; the field of
view (FOV) is $20^\circ \times 20^\circ$ for the HI-1 and $70^\circ
\times 70^\circ$ for the HI-2; the time cadence is 40 minutes for
the HI-1 and 2 hours for the HI-2 \citep{Howard2008,Eyles2009}. For
HI imaging, the transient brightness of electron-scattered sunlight
is blended with the stable brightness of dust-scattered sunlight,
star light, and planet light. Subtracted from the background
brightness, a white-light image could be discernible for the
sunlight-illuminated interplanetary transients. One or two vantage
views from HI imaging can not only recognize the background
brightness of a CIR \citep{Rouillard2008,Sheeley2008}, but also
track the speed, trajectory, and shape of interplanetary transients
\citep{J.A.Davies2009,Webb2009}. An Earth-directed MC and its
possible interaction with another MC or an incident shock could be
expected to be directly imaged in interplanetary space. The
visibility of potentially geoeffective complicated structures such
as MC-shock interaction \citep{Xiong2006a,Xiong2006b} and MC-MC
collision \citep{Xiong2007,Xiong2009,Lugaz2009} in the FOV of an HI
can give an early warning of a space weather event before its
ultimate arrival at 1 AU and hence at the Earth. On the other hand,
STElab IPS data can be employed to generate a tomographic
reconstruction of the three-dimensional large-scale solar wind
density \citep{Jackson2003}, as applied to the constraints of IPS
data from the EISCAT and MERLIN \citep{Breen2008}, and the
comparisons with density images from the Solar Mass Ejection Imager
(SMEI) \citep{Bisi2008}. When an IPS ray-path lies within the FOV of
an HI, the joint observations record simultaneous multi-scale
manifestations for interplanetary dynamics, as demonstrated in the
study of the merging between two converging CMEs on 16 May 2007
\citep{Dorrian2008}. In addition, as the optical imaging from a
single HI only records the horizontal and vertical motions of a CME,
the extra depth information could be complemented by the line of
sight (LOS) at a different perspective from another HI or from IPS.
Thus, the coordinated remote-sensing observations using both HI and
IPS cover a wide range of heliocentric distances and almost all
heliographic latitudes at any time, and record the consistent
multi-scale responses in white-light and radio wave scintillation to
the interplanetary passage of a CME.

\subsection{Observations and Models}
The interpretation of HI and IPS observational data relies
significantly on theoretical modelling, since the models provide an
insight into the underlying physical processes from its observable
manifestation in the electromagnetic spectrum. For the integral
signal along an IPS ray-path, the ambiguity of its LOS distribution
is usually limited and removed by the addition of solar corona
observations via the iterative fitting of a kinematic solar wind
model \citep{Klinglesmith1997}. This kinematic model assumes a
purely radial solar wind flow with each stream travelling at a
constant velocity, and projects the IPS ray-path onto a solar source
surface using a ballistic mapping along the spiral IMF at an
appropriate speed. By this method, the modes of solar wind occupying
different regions of the IPS ray-path can be inferred from the
distribution of dark (coronal hole) and bright (streamer) regions of
the corona onto which the projected ray-path falls
\citep{Klinglesmith1997,Bisi2006}. Similarly, the derived
distribution along the IPS ray-path constrained by the solar surface
map can be further mapped outwards along the same ballistic
trajectory to the positions of Ulysess at high latitude, and
Wind/ACE at Earth orbit, allowing a direct comparison with the
corresponding in-situ observation data \citep{Bisi2010}. The
intersecting points of ballistic trajectories mapped at different
speeds provide a qualitative indication of the location of the
compression region within a CIR \citep{Bisi2010}. Multiple
observational data at different solar distances, longitudes, and
latitudes, are assumed to be causally linked via ballistic mapping
in this kinematic solar wind model, so the IPS data could be fitted
with the extra freedom limited by other observational data
\citep{Klinglesmith1997}. For HI, the geometric influence of the
Thomson-scattering sphere significantly affects the appearance of a
CME in the FOV \citep{Vourlidas2006,Howard2009}. During the
interplanetary propagation of a CME, the relative position between
the CME and the Thomson sphere changes continuously and different
parts of the CME would be imaged by the HI at different times. For
simultaneous stereoscopic imaging, the Thomson-scattering responses
of the HI-A and HI-B could be quite asymmetrical. Moreover, the
coexistence and possible interaction between the co-rotating
background plasma in a CIR and the outwardly moving CME, can further
complicate the interpretation of white-light images. Because of the
intractable nature of analytical methods, the numerical model can
assist by providing global context and hints of what can and cannot
be observed in HI images \citep{Odstrcil2009}. Meanwhile, the
time-dependent solution from forward modelling with a numerical
heliospheric model can give the profile along any IPS ray-path.
Thus, the local IPS data are assimilated into the global numerical
model in a self-consistent way. Further, if an IPS model for the
micro-scale scattering process is logically coupled with a
heliospheric model for the macro-scale magnetohydrodynamic (MHD)
context, the synthetic observation signatures of both HI and IPS can
be derived from the multi-scale model and be linked to the
interplanetary dynamics. Motivated by the integration of
observational data as irrefutable evidence, and theoretical
modelling as a convincing interpretation, we have undertaken a
preliminary study in this paper of an incident shock propagation
within the ecliptic plane using a newly-constructed multi-scale
model. The propagation and evolution of an interplanetary CME can be
better monitored and understood by the joint efforts of an inverse
approach from the remote-sensing observations and a forward
modelling from the numerical simulations.

In this paper, the synthetic observation signatures of white-light
imaging and IPS are forwardly modelled for the propagation of an
interplanetary shock within the ecliptic plane. We present the
multi-scale numerical model in Section \ref{Sec:Model}, describe the
individual observations from the HIs in Section \ref{Sec:Optical}
and IPS in Section \ref{Sec:IPS}, analyze the correspondence and
coordination of the simultaneous white-light data and IPS data for
the same spatial position of interplanetary space in Section
\ref{Sec:Joint}, and summarize this paper and discuss the potential
of using the coordinated HI and IPS observations to identify a
possible longitudinal deflection of a CME/Shock in Section
\ref{Sec:Conclusion}.

\section{A Multi-Scale Numerical Model} \label{Sec:Model}
The observation signatures of white-light brightness and radio-wave
scintillation are physically described by our newly constructed
numerical model, an integration of a numerical MHD model for the
macro-scale driver and an IPS model for the micro-scale response.
This multi-scale model is implemented in the following steps to
causally couple the multiple physical processes at dramatically
different temporal and spatial scales.

The global MHD model for interplanetary space, governed by a set of
ideal MHD equations \citep[c.f.][]{Hu1998}, is numerically solved by
a mathematical shock-capturing algorithm of Piecewise Parabolic
Method with a Lagrangian remap (PPMLR) \citep{Colella1984,Hu2007}.
The ecliptic plane from 0.1 to 1 AU is discretized into a mesh of
200 $\times$ 360 grids with its radial spacing 0.0045 AU and
longitudinal spacing $1^\circ$. The background solar wind is
prescribed by the inner boundary conditions at 0.1 AU: number
density of $8.2 \times 10^{-4}$ m$^{-3}$, magnetic field of 250 nT,
solar rotation speed of $2.9 \times 10^{-6}$ rad s$^{-1}$, a
temperature of $4.65 \times 10^5$ K, and a bulk-flow speed of 375 km
s$^{-1}$. In this model, the background solar wind is uniform along
the longitudinal direction. Then, the initial equilibrium of
interplanetary space is disturbed by a fast MHD shock from solar
eruption, characterized by the shock nose at longitude of
$45^\circ$, the shock front width of $30^\circ$, the initial shock
speed $1150$ km s$^{-1}$, the total pressure ratio of 18 across the
front discontinuity, the disturbance duration of 2 hours. The
introduction of this shock into the simulation domain is numerically
realized by the modification of the inner boundary conditions at 0.1
AU \citep{Xiong2006a,Xiong2006b}. Thus, the propagation and
evolution of a shock wave through interplanetary space is
quantitatively described in a macro-scale by the MHD model.

The synthetic white-light imaging is generated from MHD simulation
data by the well-established Thomson-scattering theory
\citep{Vourlidas2006,Howard2009}. The location of the maximum
scattered sunlight lies on the Thomson-scattering sphere which is
centred half-way between the Sun and the observer with its diameter
equal to the Sun-observer distance. The geometry factor of the
Thomson-scattering sphere should be included to correctly interpret
the heliospheric imaging \citep{Vourlidas2006,Howard2009}. Both the
total electron number and its position relative to the
Thomson-scattering sphere significantly affect the brightness in
interplanetary space. As pointed out by \citet{Howard2009}, (1)
though the Thomson scattering itself is minimized on the Thomson
surface, the scattered sunlight is maximized on the Thomson surface;
(2) the scattered sunlight is maximized simply because it is at the
point along any LOS that is closest to the Sun, where the incident
sunlight and density are greatest; (3) the scattered intensity
becomes more spread out with distance from the Thomson surface. So
the importance of the Thomson surface in the \citet{Vourlidas2006}
method is somewhat de-emphasized in the \citet{Howard2009} method.
However, in this paper, only \citet{Vourlidas2006} method is adopted
for simplicity. The more sophisticated method of \citet{Howard2009}
will be considered in our model in the future. For an interplanetary
electron scattering photospheric light, the geometry sketch is
illustrated by \citet[Figure 1]{Vourlidas2006}, the
Thomson-scattering algorithm is available as an IDL procedure
``eltheory.pro" in the solar-soft library under the SOHO/LASCO
directory. Hence, the propagation of interplanetary disturbances in
our model directly provides their manifestation in the synthetic
white-light images.

The IPS model for the local intensity modulation of radio waves
makes use of a Born approximation to the general weak-scattering
theory \citep{Tatarski1993,Fallows2001}. The spatial fluctuation of
the local density irregularities at a micro-scale of about 200 km is
conveyed by the ambient solar wind flow, and consequently introduces
a scintillation pattern in the ($x$, $y$) plane perpendicular to the
IPS ray-path along the $z$ direction. Here ($x$, $y$, $z$) is a
cartesian coordinate system centred on the Earth. The amount of
scintillation introduced by density irregularities in the solar wind
varies according to the square of density fluctuations $\delta
N_e^2$ \citep{Salpeter1967}. As the variation of $\delta N_e$ with
heliocentric distance is not well determined from observations, it
is a common practice in IPS studies to assume that $\delta N_e$
varies as $N_e$, as suggested by \citet{Houminer1972}. The intensity
$P_{\Delta I}$ in a total spectrum of the spatial wave vector
$(k_x,k_y)$ is merely a linear superposition of all contributions
from every thin scattering layer along the IPS ray-path connecting
the Earth to a remote radio source, as described by $P_{\Delta
I}(k_x,k_y) = \int P'_{\Delta I}(k_x,k_y,z) dz$
\citep{Klinglesmith1997}. For each scattering layer with its
thickness $dz$ at distance $z$, a linear relation between the radio
intensity scintillation $P'_{\Delta I}(k_x,k_y,z)$ and the spectrum
of electron density irregularities, $P_{N_e}$, is given by the
following set of mathematical expressions with the notations of the
electron number density $N_e$, the classic electron radius $r_e =
2.82 \times 10^{-5}$ m, the observing wavelength $\lambda$, the
spatial wave number $k$, the Fresnel radius $r_f$, the spectral
exponent $\alpha$, the dissipation-associated inner scale of the
sharp drop of spectral power $k_c$, the wave vector parallel
(perpendicular) to the magnetic field $k_\parallel$ ($k_\perp$), and
the axial ratio of anisotropy degree $AR$ at a heliocentric distance
$r$ with its reference of $AR_0=8$ at $r_0=5$ solar radii
\citep{Coles1989,Klinglesmith1997,Massey1998}:
\begin{eqnarray}
 P'_{\Delta I}(k_x,k_y,z) & = & 8 \pi (r_e \lambda)^2
\sin^2(\frac{|k|^2
\lambda z}{4 \pi}) P_{N_e}(k_x,k_y,k_z=0,z) dz  \label{Equ:Pk} \\
 r_f &=& \sqrt{\lambda z}  \label{Equ:Fresnel} \\
 P_{N_e} (k_x, k_y, k_z, z)  & \varpropto & P_{N_e}
(k_{\parallel}, k_{\perp}) \varpropto N_e^2
\frac{e^{-(|k|/k_c)^2}}{(k^2_\parallel + k^2_\perp /
AR^2)^{\alpha/2}} \label{Equ:PNe}\\
AR &=& (AR_0 - 1) \cdot (r_0 / r)^{1.3} + 1 \label{Equ:AR}
\end{eqnarray}
The cylindrically-symmetric coordinate system ($k_\parallel,
k_\perp$) with its axis along the local magnetic field is
transferred from the coordinate system ($x, y, z$) defined with
respect to the IPS ray-path. In equation (\ref{Equ:Pk}), the term
$\sin^2(\frac{|k|^2 \lambda z}{4 \pi})$ is a high pass Fresnel
filter with its Fresnel radius $r_f = \sqrt{\lambda z}$. Such a
radius $r_f$ determines the maximum scale of the irregularities, at
which the amplitude fluctuation can be received at the Earth. When
the spatial spectrum of intensity $P_{\Delta I}(k_x,k_y)$ carried by
an anti-sunward speed $\mathbf{V}$ is drifted across an IPS ray-path
with its intersection angle ($90^\circ - \theta$), only the speed
component perpendicular to the IPS ray-path
$|\mathbf{V}_{\mbox{\tiny drift}}| = |\mathbf{V}| \cdot \cos\theta$
is detectable by a terrestrial radio antenna. The geometry factor
$\cos\theta$ varies along the entire IPS ray-path. Moreover, the
nearly identical diffraction pattern of radio signatures can be
sequentially received by two fixed terrestrial antennas separated by
a baseline $\mathbf{b}$. If the baseline $\mathbf{b}$ on the Earth
is approximately parallel to the drifting direction
$\mathbf{V}_{\mbox{\tiny drift}}$ of micro-scale density
irregularities in interplanetary space, the scintillation patterns
at the two telescopes will be correlated with some time lag $\tau$.
At any scattering layer $z$, the spatial correlation function
between two antennas $R'_{12}(\mathbf{b}, \tau, z) =
R'_{12}(\mathbf{S} = \mathbf{b} - \mathbf{V}_{\mbox{\tiny drift}}
\cdot \tau, z)$ can be converted from the spatial spectrum
$P'_{\Delta I}(k_x,k_y,z)$ via a Fourier transform. Specifically,
the spatial-to-temporal conversion is merely a cut in the spatial
correlation function along the direction of $\mathbf{b} -
\mathbf{V}_{\mbox{\tiny drift}} \cdot \tau$. When the baseline
$\mathbf{b}$ is zero, the Cross-Correlation Function (CCF) between
two antennas is degenerated to the Auto-Correlation Function (ACF)
for either antenna. For a single scattering layer, the CCF is simply
derived by shifting the ACF at a time lag $|\mathbf{b}| /
|\mathbf{V}_{\mbox{\tiny drift}}|$. Inversely, the drifting speed
$\mathbf{V}_{\mbox{\tiny drift}}$ can be inferred from the time lag
$\tau$, which is the principle of IPS estimation for solar wind
outflow. The drift speed in the CCF of IPS signals is the
manifestation of the flow speed of the local density irregularities.
As the amplitude of Alfv\'{e}n turbulence is much smaller in
interplanetary space compared with the inner solar corona, IPS speed
is likely to be close to the bulk plasma flow speed, at least in the
slow solar wind \citep{Klinglesmith1997}. In the fast solar wind
close to the Sun, IPS speeds may overestimate solar wind speeds
\citep{Klinglesmith1997}. However, as a remote-sensing technique, an
IPS signature involves the contributions from all of the scattering
layers along its IPS ray-path. The ability to resolve the
distribution of solar wind velocities from the cross-correlation of
IPS measurements at two sites depends upon the differences in
time-lags between the maxima in cross-correlation produced by
different streams, and thus on the baseline length $\mathbf{b}$ and
the Fresnel radius $r_f$ (Equation \ref{Equ:Fresnel}). In general,
longer baselines improve the ability to resolve streams of different
velocities \citep{Breen2008,Bisi2010}, provided that the time
interval is long enough to derive well-defined scintillation
spectra, when the observing geometry is suitable for
cross-correlation analysis. In our IPS model, the observing
wavelength (frequency) is 21 cm (1420 MHz) and the baseline
$\mathbf{b}$ is 2000 km. For the radio wave at 1420 MHz, the maximum
scale of the irregularity from a scattering layer at a distance of
$z=$ 1 AU is 177 km. As the local IPS signature is influenced by the
global parameters of bulk-flow speed, electron density, and
magnetic-field orientation, synthetic IPS data can be hierarchically
generated from MHD simulation data.

\section{White-Light Imaging} \label{Sec:Optical}
\begin{figure}
\noindent
  \includegraphics[width=0.99\textwidth]{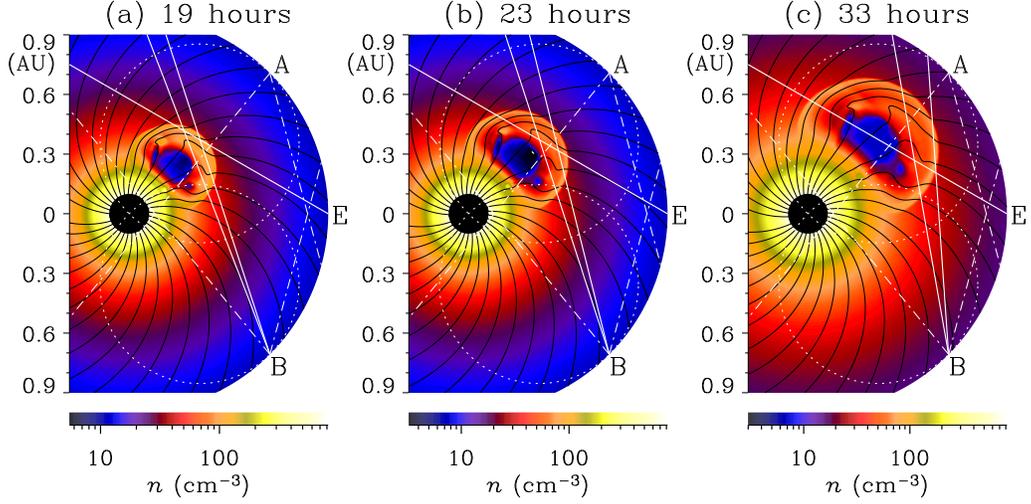}
\caption{Density distribution $n$ in the ecliptic plane for the
propagation of an interplanetary shock at (a) 19, (b) 23, and (c) 33
hours, following its initiation, with the interplanetary
magnetic-field lines shown as solid black lines. The synthetic
STEREO HI-A (HI-B) is at point `A' (`B'), with its
Thomson-scattering sphere marked by a white dotted circle, and field
of view marked by two dash-dotted lines. Note that the synthetic
STEREO A and B spacecraft are located at $45^\circ$ in the longitude
west and east from the Earth at point `E', respectively. One solid
line from the point `E' denotes an IPS ray-path at the Earth with
its elongation of $30^\circ$. Two solid lines from the point `B'
represent the elongation boundaries of relative-brightness
enhancement $(I - I_0)/I_0 \geq 0.4$ in Figure \ref{Fig:HI}.}
\label{Fig:n2D}
\end{figure}

The propagation of an interplanetary shock can be continuously
tracked at a macro-scale by white-light imaging of the inner
heliosphere. Synthetic brightness images of interplanetary
disturbances can be generated from a numerical MHD model using the
Thomson-scattering principle. In our model, an incident shock
initially launched $45^\circ$ west of the Sun-Earth line, is
characterized by a longitudinal width of $30^\circ$ along its front
and a speed of 1150 km s$^{-1}$ at its nose. The time-series
evolution of the shock is shown in Figure \ref{Fig:n2D} for density
$n$ and Figure \ref{Fig:V2D} for radial speed $v_r$. A noticeable
trailing cavity with low density is formed and expands as the shock
front propagates out from the Sun. Between this so-called density
cavity and the shock front lies the sheath region. Across the shock
front towards the sheath, the spiral IMF lines are significantly
distorted and compressed, and the bulk flow speed $v_r$, number
density $n$, plasma temperature are abruptly enhanced. Hence the
sheath downstream of a shock front is a readily observable target in
interplanetary space. White-light images are simultaneously
simulated for the twin HIs at points ``A" and ``B" in Figure
\ref{Fig:n2D}. With the longitude of $45^\circ$ beside the Earth at
1 AU, any HI has the FOV from $6^\circ$ to $60^\circ$ in the
elongation. From the combined views of HI-A and HI-B, the Sun-Earth
line is completely covered, and interplanetary space is routinely
monitored. However, the sensitivity of HI for a remote plasma parcel
depends on not only its innate electron number, but also its
heliographic position. Such a geometry dependence for white-light
imaging arises from the Thomson-scattering process, the working
principle of the HI instrument. The Thomson-scattering effect is
strongest at the sphere marked by a dotted white circle in Figure
\ref{Fig:n2D}, when the incoming direction of an incident
photospheric photon is perpendicular to the outgoing scattered light
towards the receiver. For the shock studied in this paper, the
Thomson-scattering effect is very weak for the white light received
at HI-A, as the shock propagates along the diameter of Thomson
sphere of HI-A. Moreover, only the western flank of the shock front
is within the HI-A's FOV, and the distance between this shock flank
and Thomson sphere is large due to the finite front width. In
addition, because the LOS from HI-A penetrates both the sheath with
high density and the trailing cavity with low density, the integral
signal of remote sensing can only give an average effect. Therefore,
the brightness of this shock in HI-A's FOV is very faint, as shown
in Figure \ref{Fig:HI}. The two-dimensional time-elongation image is
assembled from a series of one-dimensional slices taken at different
snapshots within the ecliptic plane. This time-elongation format,
widely used in the analysis of real observational data, has been
shown to be very effective at revealing the evolution of solar
ejecta from white-light imaging \citep{C.J.Davis2009}. Moreover, as
the background brightness $I_0$ abruptly decreases away from the
Sun, the relative-brightness enhancement $(I - I_0)/I_0$ is adopted
to highlight the brightness deviation from the initial steady state.
The relative brightness in the HI-A's FOV is far less than 0.1 until
the shock front approaches and crosses the Thomson sphere near 1 AU.
Hence the shock, heading towards the HI-A, is essentially invisible
within the HI-A's FOV. As a dramatic contrast, a bright diagonal
streak is conspicuous in HI-B data with its relative brightness up
to 0.9. This bright streak is immediately followed by a dark streak
with a sharp boundary between them. The bright and dark features in
Figure \ref{Fig:HI}b correspond to the relative positions of the
sheath and cavity in the HI-B FOV in Figure \ref{Fig:n2D}. The
sheath and cavity are separately imaged via different LOSs from
HI-B. The shock nose and the eastern flank are continuously detected
by one varying narrow LOS band from the HI-B, whose elongation is
$24^\circ \sim 27^\circ$ at 19 hours, $28^\circ \sim 32^\circ$ at 23
hours, and $35^\circ \sim 41^\circ$ at 33 hours. The asymmetry
between the white-light images of the HI-A and HI-B results from the
initial deviation of shock propagation from the Sun-Earth line.

With some assumptions or modelling, the spatial position of a local
high-density structure could be inversely inferred from its
contribution of the Thomson-scattering emission to a brightness
feature in the white-light images. Once the Thomson-scattering
source is pinpointed, the total electron number inside the source
region can be accurately calculated from the brightness corrected by
the geometry factor of the Thomson-scattering sphere. Without an
additional observation limitation, the intersection point between
the Thomson sphere and a LOS may be simply assumed to be the source
for the corresponding bright pixel in the white-light image. This
rough assumption of an interplanetary scattering source lying on the
Thomson sphere suffers from a significant underestimation of total
electron number. The underestimation degree has been studied for the
radial propagation of a single electron at various longitudes
\citep[Figure 5]{Vourlidas2006}. Directly generating the behavior of
a single propagating electron to a CME, \citet{Vourlidas2006} found
from their qualitative model that the mass underestimation (1)
exceeds a factor of 2 for a limb CME at elongations larger than
$60^\circ$, (2) also exceeds a factor of 2 for a halo CME at
elongations smaller than $20^\circ$, and (3) is never off by more
than 20\% for a CME, propagating along intermediate longitudes
($\sim 40^\circ$), even at extreme elongations. As a contrast to the
results from heliospheric imaging, the mass underestimation from the
coronagraph for heliocentric distances of less than 30 solar radii
is less than 50\%, even when a simpler assumption is made that the
electron source is solely at the so-named plane of sky perpendicular
to the Sun-observer line \citep{Vourlidas2000}. Moreover, the real
three-dimensional density distribution is more complex for an
interplanetary propagating CME against the background of the ambient
bimodal solar wind, which is difficult to retrieve from white-light
imaging. A more realistic investigation has to resort to numerical
simulations, particularly for the explanation of a practical
observation event. Though the outline of an interplanetary CME in
the white-light image can be easily identified by the excess
brightness of CME images subtracted from a pre-event background
image, the conversion from such an excess brightness to the actual
mass is not straightforward. Simultaneous imaging from two vantage
points of the twin STEREO spacecraft can significantly improve the
capability of identifying the spatial location of electron source
for an Earth-directed halo CME. However, when a front-side CME
propagates off the Sun-Earth line and fully enters the Thomson
sphere of one HI, only the other HI can discern the CME-driven shock
front, as demonstrated in Figures \ref{Fig:n2D} and \ref{Fig:HI} of
this paper. Therefore, in order to reduce the ambiguity of
interpretation for the observation data of STEREO HI-A and HI-B,
some additional observation techniques are necessary to provide some
further observational constraints.

\section{Interplanetary Scintillation Signal} \label{Sec:IPS}
Measurements of radio scintillation can provide a way of probing the
physical processes at a micro-scale of $\sim$ 200 km, such as the
electron density, bulk flow speed, magnetic-field direction, and
level of Alfv\'{e}nic turbulence. Because an IPS signal is roughly
proportional to the square of electron density (equation
\ref{Equ:PNe}) and an HI signal depends linearly on the electron
density, the IPS signal is even more sensitive to high density
regions of solar wind. When the tiny source of a strong IPS signal
is located in the bright domain of HI imaging, the IPS signal is
then confirmed to be the micro-scale manifestation of a macro-scale
interplanetary transient. Thus, the HI and IPS data can be
correlated and complement each other.

\begin{figure}
\noindent
   \includegraphics[width=0.99\textwidth]{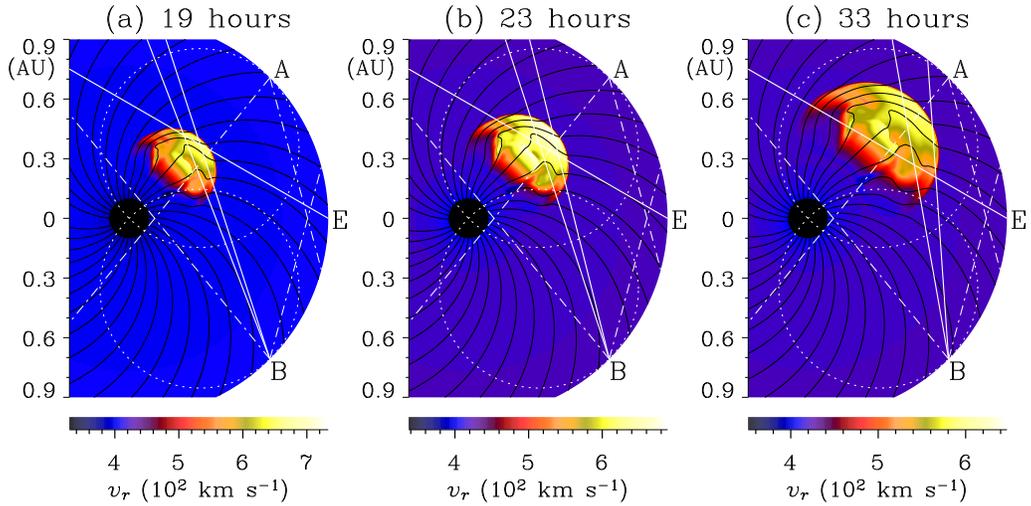}
\caption{Radial flow speed distribution $v_r$ in the ecliptic plane
for the propagation of an interplanetary shock at (a) 19, (b) 23,
and (c) 33 hours, following its initiation.} \label{Fig:V2D}
\end{figure}

\begin{figure}
\noindent
 \includegraphics[width=0.95\textwidth]{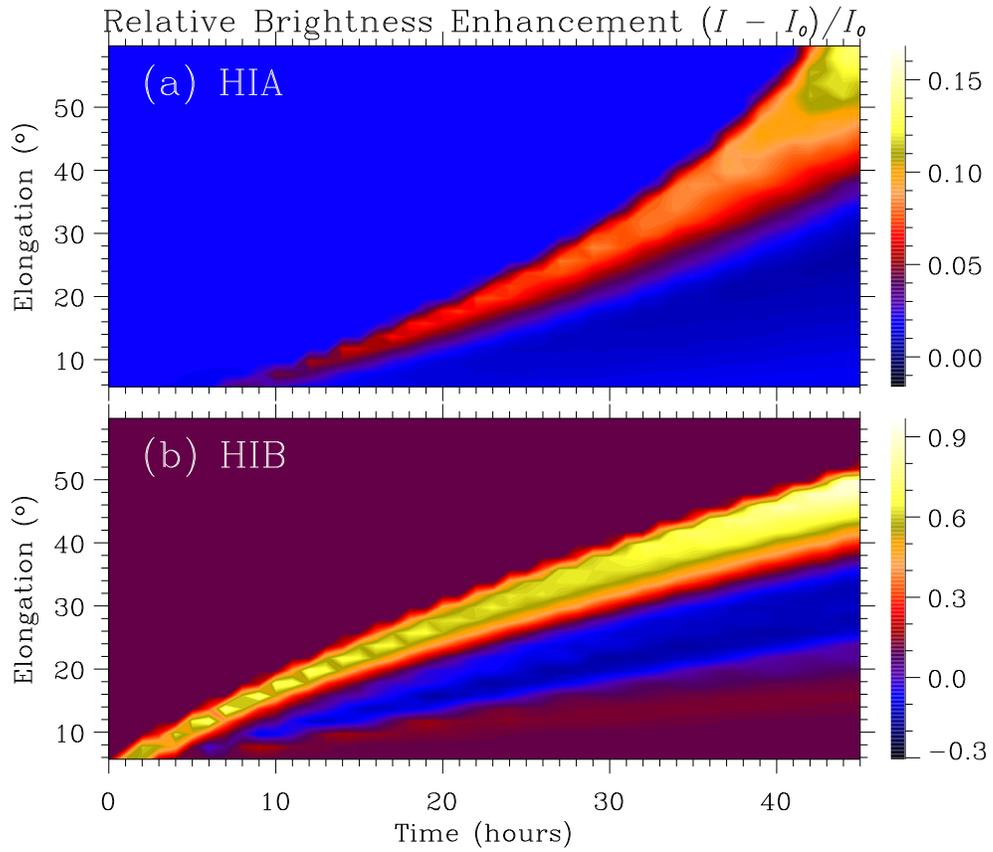}
\caption{Relative-brightness enhancement $(I-I_0)/I_0$ generated by
the time-series assembling of an elongation slice within the
ecliptic plane, according to synthetic data from the (a) HI-A and
(b) HI-B. Here, $I_0$ refers to the background
brightness.}\label{Fig:HI}
\end{figure}

\begin{figure}
\noindent
  \includegraphics[width=0.52\textwidth]{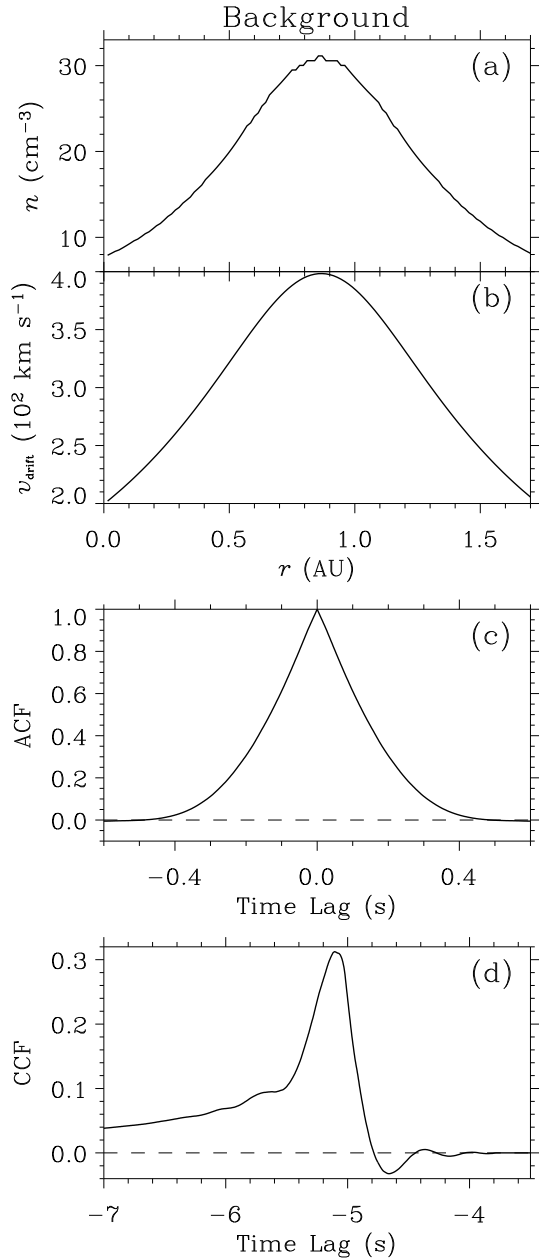}
\caption{The density $n$ (a) and speed $v_{\mbox{\tiny drift}}$ (b)
profiles along an IPS ray-path for the background solar wind. Here,
$v_{\mbox{\tiny drift}}$ refers to the speed component perpendicular
to the IPS ray-path. Below are (c) the Auto-Correlation Function
(ACF) and (d) the Cross-Correlation Function (CCF) from the IPS
observations. The baseline between the two radio antennas is 2000
km, aligned with the solar wind outflow.}\label{Fig:IPS-bg}
\end{figure}

\begin{figure}
\noindent
    \includegraphics[width=0.99\textwidth]{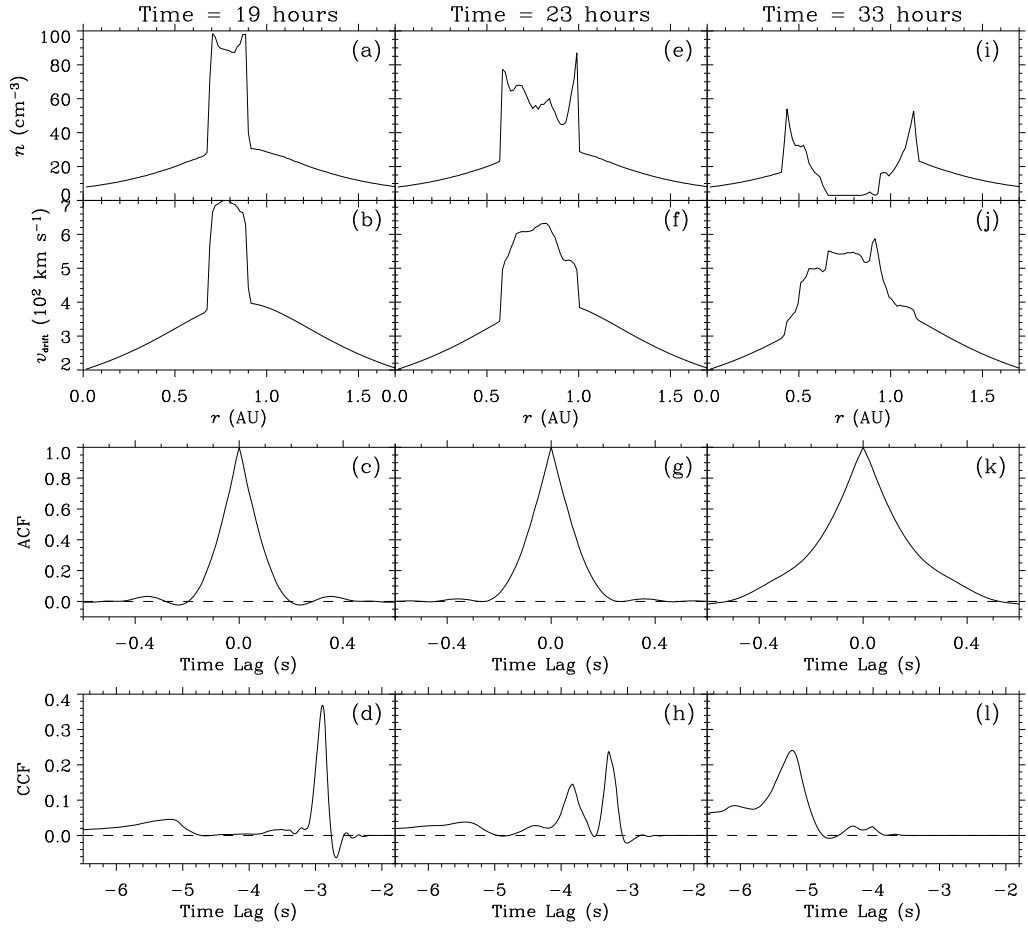}
\caption{The density $n$ (a, e, i) and speed $v_{\mbox{\tiny
drift}}$ (b, f, j) profiles along an IPS ray-path for a shock
propagation. Below are (c, g, k) the Auto-Correlation Function (ACF)
and (d, h, l) the Cross-Correlation Function (CCF) from the IPS
observations. From the left to right, the three columns refer to
three snapshots at 19, 23, and 33 hours, respectively, following the
initiation of the event.}\label{Fig:IPS-Shock}
\end{figure}

In this model, the synthetic IPS data are generated from the
distribution of the MHD data along the IPS ray-path of elongation
$30^\circ$. For the aforementioned incident shock, its front,
sheath, and cavity subsequently cross the IPS ray-path at 19, 23,
and 33 hours respectively (Figures \ref{Fig:n2D} and \ref{Fig:V2D}).
The response of radio scintillation to the shock passage is shown in
Figure \ref{Fig:IPS-Shock} and compared with the background state in
Figure \ref{Fig:IPS-bg} to highlight the differences between them.
For the ACF at 19 hours (Figure \ref{Fig:IPS-Shock}c), the negative
dip occurs as a result of oscillation of the ACF beside its central
peak. By contrast to the ACF for the background solar wind in Figure
\ref{Fig:IPS-bg}c, the angle between the IPS ray-path and the local
IMF line is found to be changed from $0^\circ$ to almost $90^\circ$.
The rotation of IMF lines in the scattering source of an IPS signal
generally means the passage of a CME across the IPS ray-path
\citep{Dorrian2008}, consistent with the global magnetic-field
configuration from the MHD model (Figure \ref{Fig:n2D}a). In
addition, the relatively small amplitude of the negative dip is
ascribed to a small axial ratio in the micro-scale interplanetary
irregularities (equation \ref{Equ:AR}), as the anisotropic
distribution of density irregularities with respect to the
magnetic-field line is dramatically reduced between the corona and
interplanetary space \citep{Armstrong1990,Grall1997}. Further, the
multiple streams across the IPS ray-path could be recorded as
corresponding multiple peaks in the CCF. Given an IPS baseline
parallel to the solar wind flow direction and long enough to
separate multiple peaks in the CCF, the time lag for each peak in
the CCF can be read to infer the flow speed of its corresponding
stream. As an instant response to the arrival of the shock front,
the negative bay in the CCF is obviously intensified, and the time
lag of the CCF is reduced from 5.1 seconds (Figure
\ref{Fig:IPS-bg}d) to 2.9 seconds (Figure \ref{Fig:IPS-Shock}d).
With an IPS baseline of 2000 km, the bulk flow speed perpendicular
to the IPS ray-path $v_t$ can be calculated. Across the shock front,
the flow speed $v_t$ is abruptly increased from 392 to 690 km
s$^{-1}$. The shock strongly disturbs the interplanetary medium as
it passes by. At 23 hours, double peaks appear in the CCF. Their
amplitudes are 0.24 at $-3.3$ seconds and 0.14 at $-3.8$ seconds in
Figure \ref{Fig:IPS-Shock}h, far less than the previous amplitude of
a single peak of 0.37 in Figure \ref{Fig:IPS-Shock}d. Two distinct
flows with $v_t=$ 606 km s$^{-1}$ and 526 km s$^{-1}$ (Figure
\ref{Fig:IPS-Shock}h) coexist along the IPS ray-path, which
correspond to the sheath and flank of this shock (Figure
\ref{Fig:n2D}b, \ref{Fig:V2D}b, \ref{Fig:IPS-Shock}e, and
\ref{Fig:IPS-Shock}f). Though the shock flank occupies a smaller
section along the IPS ray-path, it has a higher density. In other
words, the smaller number of scattering layers is largely offset by
the stronger scintillation level in each scattering layer. The total
scintillation signal from the flank is comparable to that of the
sheath. As the shock moves on, the sheath is replaced by the cavity
along the IPS ray-path (Figure \ref{Fig:n2D}c and \ref{Fig:V2D}c).
However, the cavity with very low density contributes little to the
IPS signal, and is hence ignored. Only the flank with $v_t= 385$ km
s$^{-1}$ could be captured by the CCF at the time lag of 5.2 seconds
(Figure \ref{Fig:IPS-Shock}l). The feedback of IPS measurement
(Figure \ref{Fig:IPS-Shock}) to the HI imaging (Figure \ref{Fig:HI})
could discern from which depth of HI LOS the major brightness comes.

\section{Coordinated Observations of Heliospheric Imaging and Interplanetary
Scintillation} \label{Sec:Joint} The continuous increase of
elongation for a bright pattern in an HI image is the manifestation
of a shock front driven by an outwardly propagating CME from the
Sun. The movement of the bright front is at the fast shock speed
$v_{\mbox{\tiny shock}}$. A fast shock is formed as a result of
intersecting of characteristic lines of fast magnetosonic wave $v_r
+ c_f$ upstream and downstream of the wave front. The fast shock,
behaving as a sharp discontinuity, is faster than the bulk flow
speed $v_r$, and slower than the fast magnetosonic wave $v_r + c_f$
just downstream of its front \citep{Jeffrey1964}. For a case of the
incident shock in this paper, these characteristic speeds just
downstream of shock front are shown in Figure \ref{Fig:align}b. The
furthest point within the shock nose from the Sun, defined as a
shock aphelion, is continuously tracked and presented in Figure
\ref{Fig:align}a. The slope of the shock aphelion in Figure
\ref{Fig:align}a is the shock speed $v_{\mbox{\tiny shock}}$ in
Figure \ref{Fig:align}b. Bounded by the bulk flow speed $v_r$ and
fast magnetosonic speed $v_r + c_f$, the shock speed $v_{\mbox{\tiny
shock}}$ is gradually decreased from 1150 km s$^{-1}$ at 0.1 AU to
670 km s$^{-1}$ at 1 AU during the transiting time of 45 hours. As
demonstrated in Figure \ref{Fig:align}b, the bulk flow speed $v_r$
is far greater than the fast wave speed $c_f$ because of the
supersonic solar-wind flow and the radially-decreased IMF strength.
Thus, the bulk flow speed $v_r$ just downstream of a shock front is
a reasonable approximation to the true shock speed in interplanetary
space, as demonstrated in this numerical case with the
underestimation being less than 10\%. Within observational accuracy,
there would generally be a speed match between a global bright front
in a white-light image and a local density irregularity in an IPS
signal, if the IPS ray-path lies within the FOV of the white-light
imaging. For instance, on 16 May 2007, two converging CMEs were
merged to form a discernible front in the STEREO HI-A observations,
whose speeds in the plane of sky were 325 km s$^{-1}$ and 550 km
s$^{-1}$ from the HI image, and $420 \pm 10$  km s$^{-1}$ and $520
\pm 20$ km s$^{-1}$ from IPS \citep{Dorrian2008}. The speed
agreement supports the coincidence of white-light data and IPS data
for the same solar eruption event.

\begin{figure}
\noindent
  \includegraphics[width=0.99\textwidth]{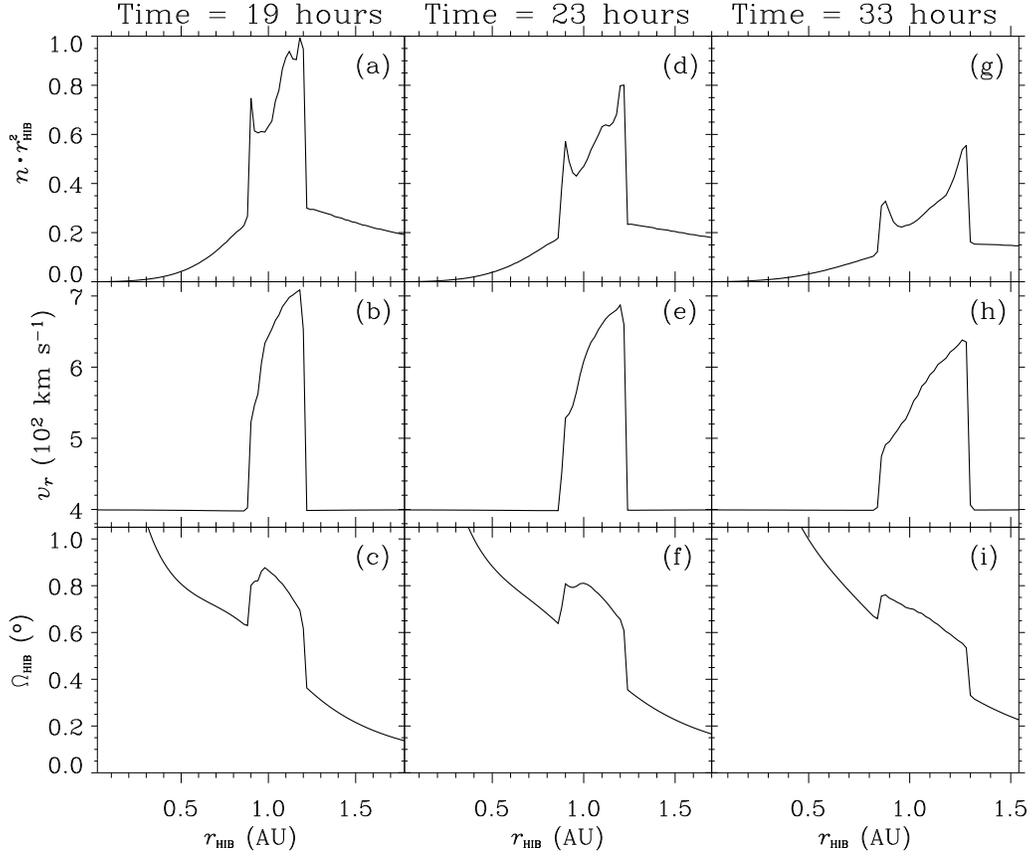}
\caption{The profiles of the electron number $n \cdot
r^2_{\mbox{\tiny HIB}}$ (a, d, g), radial bulk flow speed $v_r$ (b,
e, h), and angular speed relative to the HI-B point
$\Omega_{\mbox{\tiny HIB}}$ (c, f, i) along a specific ray-path
$r_{\mbox{\tiny HIB}}$ at the elongation of $26^\circ$ at 19 hours,
$30^\circ$ at 23 hours, and $38^\circ$ at 33 hours, viewed by the
synthetic STEREO HI-B.}\label{Fig:HI-LOS}
\end{figure}

\begin{figure}
\noindent
  \includegraphics[width=0.7\textwidth]{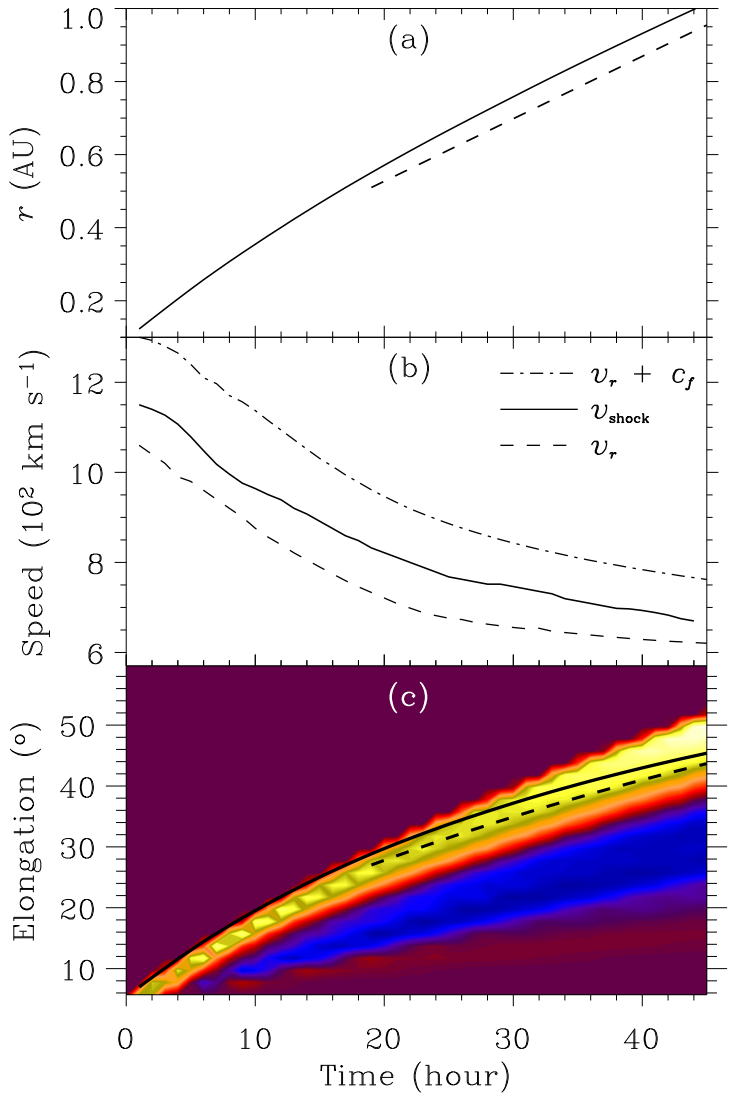}
\caption{The time-dependent variances of (a) the radial aphelion $r$
and (b) the propagation speed $v_{\mbox{\tiny shock}}$ of a shock,
as well as its manifestation in (c) the elongation of HI-B's field
of view, shown as black solid lines. In panel (b), the bulk flow
speed $v_r$ and the fast magnetosonic speed $v_r + c_f$ just
downstream of the shock front are plotted as a dashed line and a
dash-dotted line respectively. As a specific plasma parcel on the
shock front is firstly detected as an IPS irregularity source at 19
hours, its later appearances in the radial distance from the Sun (a
dashed line in panel a) and the elongation from the HI (a dashed
line in panel c) could be estimated from the IPS observation at 19
hours together with two assumptions of radial direction and constant
flow speed for the shock propagation. }\label{Fig:align}
\end{figure}

With a simultaneous observation of IPS as an additional limit, the
three-dimensional anti-sunward movement of a shock front could be
quantitatively linked to its manifestation as an outwardly-moving
bright front in a two-dimensional white-light image. Continuous
white-light imaging presented in a time-elongation format (Figure
\ref{Fig:align}c) can give the receiver-associated angular speed
$\Omega$ of the bright pattern. However, the angular speed $\Omega$
for each plasma parcel along the ray-path of each LOS is quite
different. As a result, the plasma parcels imaged earlier by one LOS
would be cut later by a series of adjoining LOS rays. The profile of
angular speed $\Omega$ along a LOS is unknown from the practical
optical imaging, as remote sensing only gives the final integral
effect along each LOS. As a contrast, the profile of various
parameters along every LOS is available from a numerical model. For
the numerical case of this paper, the radial speed from the Sun,
$v_r$, and the angular speed relative to the HI-B receiver,
$\Omega_{\mbox{\tiny HIB}}$, of each plasma parcel with its relative
brightness contribution, $n \cdot r_{\mbox{\tiny HIB}}^2$, are
resolved along each LOS, $r_{\mbox{\tiny HIB}}$ (Figure
\ref{Fig:HI-LOS}). As the shock is far away from the HI-B, the
maximum angular speed, $\Omega_{\mbox{\tiny HIB}}$, for the
effective brightness contribution would be shifted from the shock
nose (Figure \ref{Fig:HI-LOS}c) to the eastern shock flank (Figure
\ref{Fig:HI-LOS}i). The match of a shock aphelion in interplanetary
space to the outermost brightness elongation in an HI image only
happens at the near-Sun distance. For this numerical case, such an
elongation deviation occurs at 20 hours (Figure \ref{Fig:align}c),
corresponding to the radial distance of 0.57 AU (Figure
\ref{Fig:align}a). At 19 hours, with the shock front being initially
cut by an IPS ray-path (Figure \ref{Fig:n2D}a), one specific
component of the shock speed perpendicular to the IPS ray-path is
approximated to be 690 km s$^{-1}$ from the IPS observations (Figure
\ref{Fig:IPS-Shock}d). Under the assumption of radial propagation,
the three-dimensional shock speed is then calculated to be 711 km
s$^{-1}$ by its projection measured with the IPS signal. Because the
intensity of an IPS signal roughly depends on the square of electron
density (equation \ref{Equ:PNe}), and the background electron
density drops as result of the solar wind expansion, the IPS
scattering source is very close to the so-called ``p" point in the
literature, the closest point along the IPS ray-path to the Sun.
Further, the position of the plasma parcel detected as an IPS
scattering source could be calculated by the intersection between
the IPS ray-path from the Earth and the FOV of the most brightness
from the HI (Figure \ref{Fig:n2D}a). With the derived position and
speed, the imminent trajectory of the plasma parcel of the IPS
scattering source at 19 hours could be predicted, whose
manifestations in the radial distance from the Sun and the
elongation from the HI-B are shown as black dashed lines in Figure
\ref{Fig:align}a and \ref{Fig:align}c, respectively. In terms of the
radial distance and the elongation, this particular plasma parcel
follows the shock aphelion. Moreover, in Figure \ref{Fig:align}c,
the slope of the plasma parcel is obviously smaller than that of the
bright pattern. The lag of the plasma parcel in the elongation of
HI-B (Figure \ref{Fig:align}c) is ascribed to the relative distance
from HI-B. Located at the eastern flank of the shock front, the
plasma parcel detected by an IPS signal at 19 hours is further away
from HI-B than other parts at the eastern flank (Figures
\ref{Fig:n2D}a and \ref{Fig:V2D}a). For this plasma parcel, the
longer distance from the HI-B, $r_{\mbox{\tiny HIB}}$, slows down
the relative angular speed, $\Omega_{\mbox{\tiny HIB}}$, as
demonstrated in Figure \ref{Fig:HI-LOS}a-c. According to this
numerical case, the predictable appearance in the elongation of an
IPS-detected plasma parcel could serve as a lower limit for the
outermost elongation of an outwardly propagating bright front in the
heliospheric imaging. By the coupling of white-light imaging and IPS
signal, the interplanetary process of a CME/shock can be better
described and understood.

\section{Summaries and conclusions}\label{Sec:Conclusion}
\begin{figure}
\noindent \hspace*{-1.5cm}
 \includegraphics[width=1.28\textwidth]{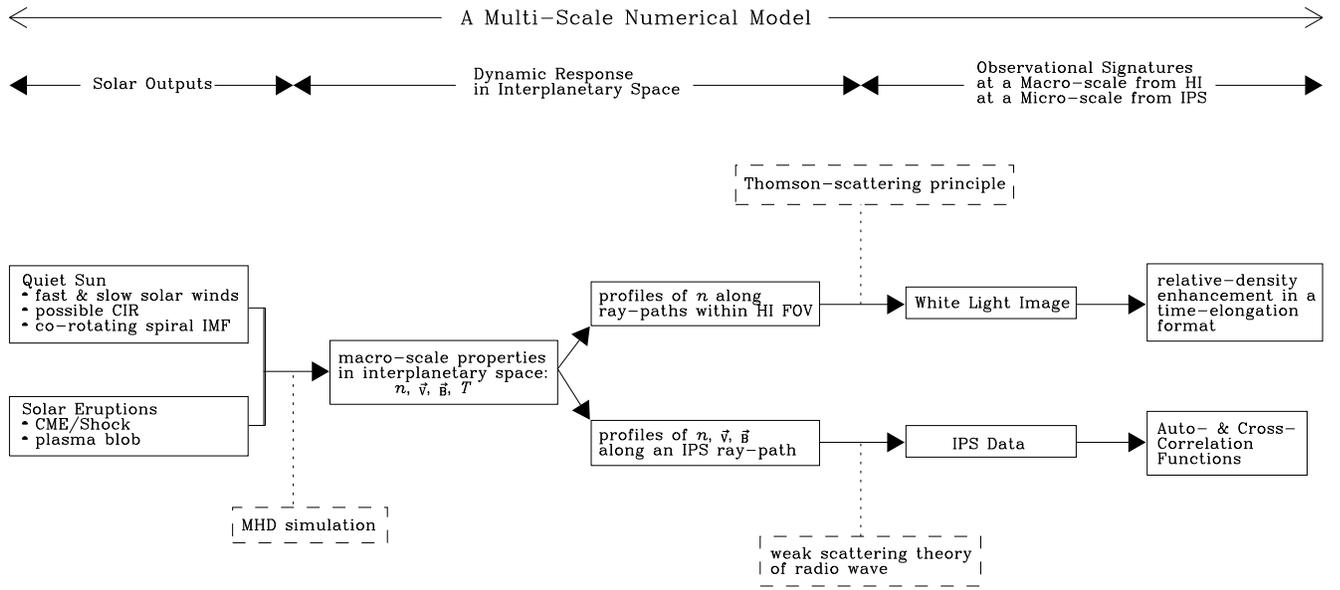}
\caption{The infrastructure of a multi-scale numerical model
directly linking interplanetary dynamics to corresponding
observational signatures from white-light imaging and IPS. As a
response to solar outputs, the macro-scale properties of electron
number density $n$, bulk-flow speed $\mathbf{V}$, magnetic field
$\mathbf{B}$, and temperature $T$ in interplanetary space are
firstly modelled from an MHD simulation. These macro-scale profiles
along any ray-path can then generate synthetic white-light image and
IPS data, using Thomson-scattering and radio-scattering theories
respectively.}\label{Fig:chart}
\end{figure}

The observational signatures of white-light imaging and IPS for the
propagation of an interplanetary shock through the ambient slow
solar wind within the ecliptic plane is analyzed via forward
modelling from a newly-constructed multi-scale numerical model. This
numerical model directly linking interplanetary dynamics to
observational signatures is summarized as a flow chart in Figure
\ref{Fig:chart}. A shock front can be sharply captured and
continuously tracked within the FOV of white-light imaging, once
being near the surface of the Thomson sphere of a receiver. The
stereoscopic imaging from two spacecraft beside the Earth can well
monitor an Earth-directed Halo CME, when the FOVs from these two
vantage points are simultaneously focused towards the Sun-Earth
line. As demonstrated by \citet{C.J.Davis2009} for a typical
Earth-directed CME on 13 December 2008, the CME viewed as a halo CME
in the coronagraph image from the Earth was symmetrically imaged by
the HI-A and HI-B onboard the two STEREO spacecraft, and was
predicted about its speed and direction at least 24 hours before its
arrival at the ACE spacecraft near 1 AU. But, if a front-side CME
propagates off the Sun-Earth line, the records of two HIs are
asymmetric. Probably, the CME is invisible to one of the HIs, if
fully entering its Thomson sphere. In this case, the
Thomson-scattering source in interplanetary space is difficult to
locate on basis of the white-light imaging from the remaining HI.
However, the ambiguity in locating the three-dimensional spatial
position from the two-dimensional bright front can be more or less
relieved with the aid of additional IPS data, if the IPS signal and
white-light imaging are coincident for the same CME event. Being cut
by an IPS ray-path, the high density-region downstream of a shock
front can be measured in terms of its bulk flow speed. When both
LOSs of HI and IPS simultaneously target the shock nose, the local
plasma parcel at the intersection point can be estimated about its
spatial position and flow speed at that time. With the assumption of
radial propagation, the plasma parcel can be predicted about its
trajectory. As the bulk flow speed just downstream of a shock front
is very near to the shock speed in interplanetary space, the
trajectory of the plasma parcel is a slower limit for the marching
shock front. Therefore, the appearance of the predicted
plasma-parcel trajectory in the HI FOV could serve as a lower limit
for the outermost elongation of an outwardly propagating bright
front in the white-light imaging.

As the most conspicuous characteristic in a white-light image, an
interplanetary brightness has multiple origins such as a shock front
and a CIR. These origins involve the compression of local plasma at
the interface between two distinct streams. The shock could be an
incident shock or a CME-driven shock. The CIR is formed as a result
of the compression between the fast and slow streams, when both
streams flow out of the rotating solar source surface at the same
heliographic latitude. As a contrast, a shock is a transient
disturbance from solar eruptions, and a CIR is an ever-changing
periodic structure in the background of interplanetary space.
Continuously imaged in white light, both a shock
\citep{J.A.Davies2009} and a CIR \citep{Rouillard2008} have the
variance and movement in their optical brightness. Sometimes, the
brightness of a CIR can be enhanced, when a preceding slow plasmoid
is firstly swept and then entrained by the following fast CIR. Such
a plasmoid imaged by an HI could be a plasma blob disconnected from
the cusp point of a coronal helmet streamer \citep{Rouillard2008} or
a small-scale MC \citep{Rouillard2009}. Furthermore, when a CIR in
the Sun-rooted spiral morphology blocks the trajectory of an
energetic CME, the collision can lead to the CME becoming entrained
by the CIR and the CIR being warped by the entrained CME. The
interplanetary dynamics of the CME-CIR interaction would be
manifested in white-light imaging as a more complex behavior of the
brightness. Meanwhile, the IPS technique has its observational
capability to probe the micro-scale density fluctuation inside the
macro-scale brightness imaged by an HI. Hence, the coordinated
remote-sensing observations of white-light and IPS are efficient to
monitor the whole interplanetary space.

The joint observations of white-light and IPS can provide the
consistent observational evidence for the possible longitudinal
deflection of a CME/shock in interplanetary space. The radial and
latitudinal movements of a CME are recorded in the two-dimensional
white-light image, once the CME is within the FOV of the HI and near
the Thomson sphere surface of the HI. The longitudinal movement of a
CME could be inferred from the continuous white-light imaging with
complementary IPS observation. The IPS signal gives the drifting
speed of local density irregularity perpendicular to the IPS
ray-path from the Earth. Considered as the global bulk flow speed,
the local IPS drifting speed derives the three-dimensional flow
speed with the assumption of radial propagation. The derived radial
flow speed can serve as a lower limit in the elongation of
white-light imaging for an outwardly propagating CME, as interpreted
in Section \ref{Sec:Joint} and demonstrated in Figure
\ref{Fig:align}c. The deviation of the predicted elongation-time
curve suggests the non-radial propagation of the CME. If the
latitudinal deflection is excluded from the HI imaging, the
non-radial propagation should come from the longitudinal direction.
The longitudinal deflection can be again confirmed by the
stereoscopic white-light imaging from the HI-A and HI-B instruments,
as shown in Figure \ref{Fig:n2D}. For instance, if an Earth-directed
CME is gradually deflected to the west and is finally enclosed by
the HI-A's Thomson sphere, the previously perfect symmetry between
the white-light images of two HIs is gradually broken, and the HI-A
image becomes darker and darker. The deflection effect clarifies the
disappearance of the CME-associated bright front in the FOV of HI-A.
In fact, the shock aphelion in this paper does deviate to the west,
because the shock front is quasi-perpendicular in the west and
quasi-parallel in the east as a result of the spiral configuration
of the IMF \citep{Hu1998}. However, the total deflection angle
during the Sun-Earth space is only $3^\circ$, and is too small to be
discerned by the observations of HI and IPS. The ignorable
longitudinal deflection in this paper is caused by the unimodal
ambient stream of slow solar wind. If a CIR is incorporated into our
model as one feature of the background, the CME deflection may be
significant due to the CME-CIR interaction \citep{Hu1998}.
Alternatively, if the initial eruptions of an early slow CME and a
late fast CME are at an appropriate angular difference, the contrary
deflections of the two CMEs could be noticeable during the
interplanetary process of oblique collision \citep{Xiong2009}. These
significant deflections are as a result of the CME-CIR interacting
or CME-CME coupling, and should be reflected from the observational
signatures of white-light and IPS. These observational signatures
will be further explored as a continuation to the preliminary
results presented in this paper.

\section{Acknowledgments}
We were supported by the Science \& Technology Facilities Council
(STFC), UK.






\bibliographystyle{elsarticle-harv}
\bibliography{mxiong}

\begin{thebibliography}{45}
\expandafter\ifx\csname natexlab\endcsname\relax\def\natexlab#1{#1}\fi
\expandafter\ifx\csname url\endcsname\relax
  \def\url#1{\texttt{#1}}\fi
\expandafter\ifx\csname urlprefix\endcsname\relax\def\urlprefix{URL }\fi

\bibitem[{Armstrong et~al.(1990)Armstrong, Coles, Rickett, and
  Kojima}]{Armstrong1990}
Armstrong, J.~W., Coles, W.~A., Rickett, B.~J., Kojima, M., 1990. Observations
  of field-aligned density fluctuations in the inner solar wind. Astrophys. J.
  358, 685--692.

\bibitem[{Bisi(2006)}]{Bisi2006}
Bisi, M.~M., 2006. Interplanetary scintillation studies of the large-scale
  structure of the solar wind. Ph.D. {Thesis}, {Aberystwyth University}, Wales,
  UK.

\bibitem[{Bisi et~al.(2010)Bisi, Fallows, Breen, and O'Neill}]{Bisi2010}
Bisi, M.~M., Fallows, R.~A., Breen, A.~R., O'Neill, I.~J., 2010. Interplanetary
  scintillation observations of stream interaction regions in the solar wind.
  Solar Phys. 261, 149--172.

\bibitem[{Bisi et~al.(2008)Bisi, Jackson, Hick, Buffington, Odstrcil, and
  Clover}]{Bisi2008}
Bisi, M.~M., Jackson, B.~V., Hick, P.~P., Buffington, A., Odstrcil, D., Clover,
  J.~M., 2008. Three-dimensional reconstructions of the early {November 2004
  Coordinated Data Analysis Workshop} geomagnetic storms: {Analyses of STELab
  IPS speed and SMEI} density data. J. Geophys. Res. 113~(52), A00A11.

\bibitem[{Breen et~al.(2008)Breen, Fallows, Bisi, Jones, Jackson, Kojima,
  Dorrian, Middleton, Thomasson, and Wannberg}]{Breen2008}
Breen, A.~R., Fallows, R.~A., Bisi, M.~M., Jones, R.~A., Jackson, B.~V.,
  Kojima, M., Dorrian, G.~D., Middleton, H.~R., Thomasson, P., Wannberg, G.,
  2008. The solar eruption of 2005 {May} 13 and its effects: {Long}-baseline
  interplanetary scintillation observations of the {Earth}-directed coronal
  mass ejection. Astrophys. J. 683, L79--L82.

\bibitem[{Burlaga et~al.(1987)Burlaga, Behannon, and Klein}]{Burlaga1987}
Burlaga, L.~F., Behannon, K.~W., Klein, L.~W., 1987. Compound streams, magnetic
  clouds, and major geomagnetic storms. J. Geophys. Res. 92~(A6), 5725--5734.

\bibitem[{Burlaga et~al.(2002)Burlaga, Plunkett, and Cyr}]{Burlaga2002}
Burlaga, L.~F., Plunkett, S.~P., Cyr, O. C.~S., 2002. Successive {CMEs} and
  complex ejecta. J. Geophys. Res. 107.

\bibitem[{Colella and Woodward(1984)}]{Colella1984}
Colella, P., Woodward, P.~R., 1984. The piecewise parabolic method {(PPM)} for
  gas-dynamical simulations. J. Comput. Phys. 54, 174--201.

\bibitem[{Coles and Harmon(1989)}]{Coles1989}
Coles, W.~A., Harmon, J.~K., 1989. Propagation observations of the solar wind
  near the {Sun}. Astrophys. J. 337, 1023--1034.

\bibitem[{Dal~Lago et~al.(2006)Dal~Lago, Gonzalez, Balmaceda, Vieira, Echer,
  Guarnieri, Santos, da~Silva, de~Lucas, de~Gonzalez, Schwenn, and
  Schuch}]{Dal_Lago2006}
Dal~Lago, A., Gonzalez, W.~D., Balmaceda, L.~A., Vieira, L. E.~A., Echer, E.,
  Guarnieri, F.~L., Santos, J., da~Silva, M.~R., de~Lucas, A., de~Gonzalez, A.
  L.~C., Schwenn, R., Schuch, N.~J., 2006. The 17-22 {October} (1999)
  solar-interplanetary-geomagnetic event: {Very} intense geomagnetic storm
  associated with a pressure balance between interplanetary coronal mass
  ejection and a high-speed stream. J. Geophys. Res. 111, A07S14.

\bibitem[{Davies et~al.(2009)Davies, Harrison, Rouillard, Jr., Perry, Bewsher,
  Davis, Eyles, Crothers, and Brown}]{J.A.Davies2009}
Davies, J.~A., Harrison, R.~A., Rouillard, A.~P., Jr., N. R.~S., Perry, C.~H.,
  Bewsher, D., Davis, C.~J., Eyles, C.~J., Crothers, S.~R., Brown, D.~S., 2009.
  A synoptic view of solar transient evolution in the inner heliosphere using
  the {Heliospheric Imagers} on {STEREO}. Geophys. Res. Lett. 36, L02102.

\bibitem[{Davis et~al.(2009)Davis, Davies, Lockwood, Rouillard, Eyles, and
  Harrison}]{C.J.Davis2009}
Davis, C.~J., Davies, J.~A., Lockwood, M., Rouillard, A.~P., Eyles, C.~J.,
  Harrison, R.~A., 2009. Sterescopic imaging of an earth-impacting solar
  coronal mass ejection: {A} major milestone for the {STEREO} mission. Geophys.
  Res. Lett. 36, L08102.

\bibitem[{Dorrian et~al.(2008)Dorrian, Breen, Brown, Davies, Fallows, and
  Rouillard}]{Dorrian2008}
Dorrian, G.~D., Breen, A.~R., Brown, D.~S., Davies, J.~A., Fallows, R.~A.,
  Rouillard, A.~P., 2008. Simultaneous interplanetary scintillation and
  {Heliospheric Imager} observations of a coronal mass ejection. Geophys. Res.
  Lett. 35, L24104.

\bibitem[{Eyles et~al.(2009)Eyles, Harrison, Davis, Waltham, Shaughnessy,
  Mapson-Menard, Bewsher, Crothers, Davies, Simnett, Howard, Moses, Newmark,
  Socker, Halain, Defise, Mazy, and Rochus}]{Eyles2009}
Eyles, C.~J., Harrison, R.~A., Davis, C.~J., Waltham, N.~R., Shaughnessy,
  B.~M., Mapson-Menard, H. C.~A., Bewsher, D., Crothers, S.~R., Davies, J.~A.,
  Simnett, G.~M., Howard, R.~A., Moses, J.~D., Newmark, J.~S., Socker, D.~G.,
  Halain, J.-P., Defise, J.-M., Mazy, E., Rochus, P., 2009. The {Heliospheric
  Imagers} onboard the {STEREO} mission. Solar Phys. 254, 387--445.

\bibitem[{Fallows(2001)}]{Fallows2001}
Fallows, R.~A., 2001. Studies of the solar wind through a solar cycle. Ph.D.
  {Thesis}, {Aberystwyth University}, Wales, UK.

\bibitem[{Fallows et~al.(2006)Fallows, Breen, Bisi, Jones, and
  Wannberg}]{Fallows2006}
Fallows, R.~A., Breen, A.~R., Bisi, M.~M., Jones, R.~A., Wannberg, G., 2006.
  Dual-frequency interplanetary scintillation observations of the solar wind.
  Geophys. Res. Lett. 33~(11), L11106.

\bibitem[{Grall et~al.(1997)Grall, Coles, Spangler, Sakurai, and
  Harmon}]{Grall1997}
Grall, R.~R., Coles, W.~A., Spangler, S.~R., Sakurai, T., Harmon, J.~K., 1997.
  Observations of field-aligned density microstructure near the sun. J.
  Geophys. Res. 102~(A1), 263--274.

\bibitem[{Hewish et~al.(1964)Hewish, Scott, and Willis}]{Hewish1964}
Hewish, A., Scott, P.~F., Willis, D., 1964. Interplanetary scintillation of
  small diameter radio sources. Nature 203, 1214.

\bibitem[{Houminer and Hewish(1972)}]{Houminer1972}
Houminer, Z., Hewish, A., 1972. Long-lived sectors of enhanced density
  irregularities in the solar wind. Planetary and Space Science 20~(10),
  1703--1716.

\bibitem[{Howard et~al.(2008)Howard, Moses, Vourlidas, Newmark, Socker,
  Plunkett, Korendyke, Cook, Hurley, Davila, Thompson, St~Cyr, Mentzell,
  Mehalick, Lemen, Wuelser, Duncan, Tarbell, Wolfson, Moore, Harrison, Waltham,
  Lang, Davis, Eyles, Mapson-Menard, Simnett, Halain, Defise, Mazy, Rochus,
  Mercier, Ravet, Delmotte, Auchere, Delaboudiniere, Bothmer, Deutsch, Wang,
  Rich, Cooper, Stephens, Maahs, Baugh, McMullin, and Carter}]{Howard2008}
Howard, R.~A., Moses, J.~D., Vourlidas, A., Newmark, J.~S., Socker, D.~G.,
  Plunkett, S.~P., Korendyke, C.~M., Cook, J.~W., Hurley, A., Davila, J.~M.,
  Thompson, W.~T., St~Cyr, O.~C., Mentzell, E., Mehalick, K., Lemen, J.~R.,
  Wuelser, J.~P., Duncan, D.~W., Tarbell, T.~D., Wolfson, C.~J., Moore, A.,
  Harrison, R.~A., Waltham, N.~R., Lang, J., Davis, C.~J., Eyles, C.~J.,
  Mapson-Menard, H., Simnett, G.~M., Halain, J.~P., Defise, J.~M., Mazy, E.,
  Rochus, P., Mercier, R., Ravet, M.~F., Delmotte, F., Auchere, F.,
  Delaboudiniere, J.~P., Bothmer, V., Deutsch, W., Wang, D., Rich, N., Cooper,
  S., Stephens, V., Maahs, G., Baugh, R., McMullin, D., Carter, T., 2008. {Sun
  Earth Connection Coronal and Heliospheric Investigation (SECCHI)}. Space Sci.
  Review 136, 67--115.

\bibitem[{Howard and Tappin(2009)}]{Howard2009}
Howard, T.~A., Tappin, S.~J., 2009. Interplanetary coronal mass ejections
  observed in the heliosphere: 1. {Review} of theory. Space Sci. Review 147,
  31--54.

\bibitem[{Hu(1998)}]{Hu1998}
Hu, Y.~Q., 1998. Asymmetric propagation of flare-generated shocks in the
  heliospheric equatorial plane. J. Geophys. Res. 103~(A7), 14,631--14,642.

\bibitem[{Hu et~al.(2007)Hu, Guo, and Wang}]{Hu2007}
Hu, Y.~Q., Guo, X.~C., Wang, C., 2007. On the ionospheric and reconnection
  potentials of the earth: {Results} from global {MHD} simulations. J. Geophys.
  Res. 112, A07215.

\bibitem[{Jackson et~al.(2003)Jackson, Hick, Buffington, Kojima, Tokumaru,
  Fujiki, Ohmi, and Yamashita}]{Jackson2003}
Jackson, B.~V., Hick, P.~P., Buffington, A., Kojima, M., Tokumaru, M., Fujiki,
  K., Ohmi, T., Yamashita, M., 2003. Time-dependent tomography of hemispheric
  features using interplanetary scintillation ({IPS}) remote-sensing
  observations. In: Velli, M., Bruno, R., Malara, F., Bucci, B. (Eds.), Solar
  Wind Ten. Vol. AIP Conf. of 679. p.~75.

\bibitem[{Jeffrey and Taniuti(1964)}]{Jeffrey1964}
Jeffrey, A., Taniuti, T., 1964. Non-Linear Wave Propagation with Application to
  Physics and Magnetohydrodynamics. Academic Press, New York.

\bibitem[{Kaiser(2008)}]{Kaiser2008}
Kaiser, M. L.;~Kucera, T. A. D. J. M. S. C. O. C. G. M. C.~E., 2008. The
  {STEREO} mission: {An} introduction. Space Sci. Review 136, 5--16.

\bibitem[{Klinglesmith(1997)}]{Klinglesmith1997}
Klinglesmith, M., 1997. The polar solar wind from 2.5 to 40 solar radii:
  {Results} of intensity scintillation measurements. Ph.D. {Thesis},
  {University} of California, San Diego, USA.

\bibitem[{Lepping et~al.(1997)Lepping, Burlaga, Szabo, Ogilvie, Mish,
  Vassiliadis, Lazarus, Steinberg, Farrugia, Janoo, and Mariani}]{Lepping1997}
Lepping, R.~P., Burlaga, L.~F., Szabo, A., Ogilvie, K.~W., Mish, W.~H.,
  Vassiliadis, D., Lazarus, A.~J., Steinberg, J.~T., Farrugia, C.~J., Janoo,
  L., Mariani, F., 1997. The {Wind} magnetic cloud and events of {October}
  18-20, 1995: {Interplanetary} properties and as triggers for geomagnetic
  activity. J. Geophys. Res. 102~(A7), 14,049--14,063.

\bibitem[{Lugaz et~al.(2009)Lugaz, Vourlidas, Roussev, and Morgan}]{Lugaz2009}
Lugaz, N., Vourlidas, A., Roussev, I.~I., Morgan, H., 2009. Solar-terrestrial
  simulation in the {STEREO} era: {The} 24-25 {January} 2007 eruptions. Solar
  Phys. 256, 269--284.

\bibitem[{Massey(1998)}]{Massey1998}
Massey, W., 1998. Measuring intensity scintillations at the {Very Long Baseline
  Array (VLBA)} to probe the solar wind near the {Sun}. Master {Thesis},
  {University} of California, San Diego, USA.

\bibitem[{McComas et~al.(2000)McComas, Barraclough, Funsten, Gosling,
  Santiago-Munoz, Skoug, Goldstein, Neugebauer, Riley, and
  Balogh}]{McComas2000}
McComas, D.~J., Barraclough, B.~L., Funsten, H.~O., Gosling, J.~T.,
  Santiago-Munoz, E., Skoug, R.~M., Goldstein, B.~E., Neugebauer, M., Riley,
  P., Balogh, A., 2000. Solar wind observations over {Ulysses'} first full
  polar orbit. J. Geophys. Res. 105, 10,419--10,434.

\bibitem[{Odstrcil and Pizzo(2009)}]{Odstrcil2009}
Odstrcil, D., Pizzo, V.~J., 2009. Numerical heliospheric simulations as
  assisting tool for interpretation of observations by {STEREO Heliospheric
  Imagers}. Solar Phys. 259, 297--309.

\bibitem[{Rouillard et~al.(2008)Rouillard, Davies, Forsyth, Rees, Davis,
  Harrison, Lockwood, Bewsher, Crothers, Eyles, Hapgood, and
  Perry}]{Rouillard2008}
Rouillard, A.~P., Davies, J.~A., Forsyth, R.~J., Rees, A., Davis, C.~J.,
  Harrison, R.~A., Lockwood, M., Bewsher, D., Crothers, S.~R., Eyles, C.~J.,
  Hapgood, M., Perry, C.~H., 2008. First imaging of corotating interaction
  regions using the {STEREO} spacecraft. Geophys. Res. Lett. 35, L10110.

\bibitem[{Rouillard et~al.(2009)Rouillard, Savani, Davies, Lavraud, Forsyth,
  Morley, Opitz, Sheeley, Burlaga, Sauvaud, Simunac, Luhmann, Galvin, Crothers,
  Davis, Harrison, Lockwood, Eyles, Bewsher, and Brown}]{Rouillard2009}
Rouillard, A.~P., Savani, N.~P., Davies, J.~A., Lavraud, B., Forsyth, R.~J.,
  Morley, S.~K., Opitz, A., Sheeley, N.~R., Burlaga, L.~F., Sauvaud, J.-A.,
  Simunac, K. D.~C., Luhmann, J.~G., Galvin, A.~B., Crothers, S.~R., Davis,
  C.~J., Harrison, R.~A., Lockwood, M., Eyles, C.~J., Bewsher, D., Brown,
  D.~S., 2009. A multispacecraft analysis of a small-scale transient entrained
  by solar wind streams. Solar Phys. 256, 307--326.

\bibitem[{Salpeter(1967)}]{Salpeter1967}
Salpeter, E.~E., 1967. Interplanetary scintillations{: I. Theory}. Astrophys.
  J. 147, 433.

\bibitem[{Sheeley et~al.(2008)Sheeley, Herbst, Palatchi, Wang, Howard, Moses,
  Vourlidas, Newmark, Socker, Plunkett, Korendyke, Burlaga, Davila, Thompson,
  St~Cyr, Harrison, Davis, Eyles, Halain, Wang, Rich, Battams, Esfandiari, and
  Stenborg}]{Sheeley2008}
Sheeley, N.~R., Herbst, A.~D., Palatchi, C.~A., Wang, Y.-M., Howard, R.~A.,
  Moses, J.~D., Vourlidas, A., Newmark, J.~S., Socker, D.~G., Plunkett, S.~P.,
  Korendyke, C.~M., Burlaga, L.~F., Davila, J.~M., Thompson, W.~T., St~Cyr,
  O.~C., Harrison, R.~A., Davis, C.~J., Eyles, C.~J., Halain, J.~P., Wang, D.,
  Rich, N.~B., Battams, K., Esfandiari, E., Stenborg, G., 2008. {SECCHI}
  observations of the {Sun's} garden-hose density spiral. Astrophys. J. 674,
  109.

\bibitem[{Tatarski et~al.(1993)Tatarski, Ishimaru, and
  Zavorotny}]{Tatarski1993}
Tatarski, V., Ishimaru, A., Zavorotny, V., 1993. Wave propagation in a random
  medium (Scintillation). SPIE Press, Bellingham.

\bibitem[{Vourlidas and Howard(2006)}]{Vourlidas2006}
Vourlidas, A., Howard, R.~A., 2006. The proper treatment of coronal mass
  ejection brightness: {A} new methodology and implications for observations.
  Astrophys. J. 642, 1216--1221.

\bibitem[{Vourlidas et~al.(2000)Vourlidas, Subramanian, Dere, and
  Howard}]{Vourlidas2000}
Vourlidas, A., Subramanian, P., Dere, K.~P., Howard, R.~A., 2000. Large-angle
  spectrometric coronagraph measurements of the energetics of coronal mass
  ejections. Astrophys. J. 534, 456--467.

\bibitem[{Webb et~al.(2009)Webb, Howard, Fry, Kuchar, Odstrcil, Jackson, Bisi,
  Harrison, Morrill, Howard, and Johnston}]{Webb2009}
Webb, D.~F., Howard, T.~A., Fry, C.~D., Kuchar, T.~A., Odstrcil, D., Jackson,
  B.~V., Bisi, M.~M., Harrison, R.~A., Morrill, J.~S., Howard, R.~A., Johnston,
  J.~C., 2009. Study of {CME} propagation in the inner heliosphere: {SOHO
  LASCO, SMEI and STEREO HI} observations of the {January} 2007 events. Solar
  Phys. 256, 239--267.

\bibitem[{Wei et~al.(2006)Wei, Feng, Yang, and Zhong}]{Wei2006}
Wei, F.~S., Feng, X.~S., Yang, F., Zhong, D., 2006. A new non-pressure-balanced
  structure in interplanetary space: Boundary layers of magnetic clouds. J.
  Geophys. Res. 111.

\bibitem[{Xiong et~al.(2009)Xiong, Zheng, and Wang}]{Xiong2009}
Xiong, M., Zheng, H.~N., Wang, S., 2009. Magnetohydrodynamic simulation of the
  interaction between two interplanetary magnetic clouds and its consequent
  geoeffectiveness: 2. {Oblique} collision. J. Geophys. Res. 114, A11101.

\bibitem[{Xiong et~al.(2006{\natexlab{a}})Xiong, Zheng, Wang, and
  Wang}]{Xiong2006a}
Xiong, M., Zheng, H.~N., Wang, Y.~M., Wang, S., 2006{\natexlab{a}}.
  Magnetohydrodynamic simulation of the interaction between interplanetary
  strong shock and magnetic cloud and its consequent geoeffectiveness. J.
  Geophys. Res. 111, A08105.

\bibitem[{Xiong et~al.(2006{\natexlab{b}})Xiong, Zheng, Wang, and
  Wang}]{Xiong2006b}
Xiong, M., Zheng, H.~N., Wang, Y.~M., Wang, S., 2006{\natexlab{b}}.
  Magnetohydrodynamic simulation of the interaction between interplanetary
  strong shock and magnetic cloud and its consequent geoeffectiveness: 2.
  {Oblique} collision. J. Geophys. Res. 111, A11102.

\bibitem[{Xiong et~al.(2007)Xiong, Zheng, Wu, Wang, and Wang}]{Xiong2007}
Xiong, M., Zheng, H.~N., Wu, S.~T., Wang, Y.~M., Wang, S., 2007.
  Magnetohydrodynamic simulation of the interaction between two interplanetary
  magnetic clouds and its consequent geoeffectiveness. J. Geophys. Res. 112,
  A11103.

\end{thebibliography}








\clearpage

\end{document}